\documentclass[twocolumn,english,prb]{revtex4-1}
\usepackage{hyperref}
\usepackage{amsmath}
\usepackage{amssymb}

\newcommand{\Tr}{\operatorname{Tr}}

\begin{document}
	
\title{Two-dimensional Coulomb glass as a model for vortex pinning in superconducting films.}

\author{I.\,Poboiko and  M.\,V.\,Feigel'man}

\address{L.D. Landau Institute for Theoretical Physics RAS, Moscow 119334, Russia\\
	Skolkovo Institute for Science and Technology, Moscow 121205, Russia\\
	National Research University <<Higher School of Economics>>, Moscow 101000, Russia}

\begin{abstract}
A glass model of vortex pinning in highly disordered thin superconducting films in magnetic fields $B \ll H_{c2}$ at low temperatures is proposed. Strong collective pinning of a vortex system realized in  disordered superconductors that are close to the quantum phase transition to the insulating phase --- such as $\mathrm{In O}_x$, $\mathrm{Nb N}$, $\mathrm{Ti N}$, $\mathrm{Mo Ge}$, nano-granular aluminium, and others --- is considered theoretically for the first time. Utilizing the replica trick developed for the spin glass theory, we demonstrate that such vortex system is in non-ergodic state of glass type with large kinetic inductance per square $L_K$. Distribution function of local pinning energies is calculated, and it is shown that it possesses a wide gap, i.e. the probability to find a weakly pinned vortex is extremely low.
\end{abstract}

\maketitle

\textit{1. Introduction.}
In this Letter  we study  strongly disordered superconducting films subject to magnetic field $B \ll H_{c2}$ at low temperatures. The main interest for this problem emerges from active experimental research in this area (see e.g. review \cite{NPhys20}; a more detailed discussion regarding some of experiments \cite{Yazdani2013,Sacepe2019} is discussed in the end of  the paper). 
The main issue we need to study is the competition between strong pinning of each individual  vortex by disorder and repulsion between vortices. Strong pinning corresponds to the energy variations of the order of vortex core energy itself when vortex is moved by distance of the order of the core size $\xi$.   Such strong pinning emerges because the order parameter itself is strongly fluctuating~\cite{AnnPhys2010}. The regular vortex lattice in such situation does not appear, and even the short range order is absent, but the vortex density is constant on average and is fixed by the external magnetic field. 
It is very important for such state to exist that the energy of shear deformations of the vortex lattice is much smaller compared to the energy of elastic deformations (according to \cite{AAA}, the energy of triangular lattice differs by just 2\% below the energy of a square lattice). 
Absence of the short range order makes the classical approach due to  A.I.Larkin~\cite{Larkin1970} (see also paper~\cite{LO1979} and reviews~\cite{PinningReview1,PinningReview2,PinningReview3}) inapplicable, since these papers treat the potential of defects as a perturbation compared to the energy of elastic deformations of a vortex lattice (in the model of weak collective pinning), or  study pinning of isolated vortices neglecting the interaction between them; both approaches are inapplicable to the problem at hands.
We also mention  theory of  strong pinning~\cite{Lab,Gesh1,Gesh2}, where strong impurities were considered, and the interaction between vortices was considered by means of  elasticity theory for the vortex lattice; it was possible due to low concentration of strong impurities. Our situation is different: the defects are strong and their concentration is high.

We  develop a theory of  vortex glass in a situation, which reminds  the ``Coulomb glass'' state realized  in the model of the Coulomb gap proposed by Efros and Shklovskii~\cite{ES}, but in a situation when the interaction between particles (vortices in our case) is logarithmic repulsion  $U(r) = U_0\ln\frac{a}{r}$, instead  of usual Coulomb one.
Here the constant is $U_0 = \frac{\Phi_0^2 d}{8\pi^2\lambda^2}$ for a thin superconducting film of thickness $d$, which is much smaller than London penetration depth $\lambda$. 
Strictly speaking, on the largest distances $r \geq \lambda_{2D} = 2\lambda^2/d$ the interaction energy is no longer logarithmic, it decays as $ \propto 1/r$; however, we will consider superconductors with a very high ratio $\lambda/d \geq 100$ (which is easily realizable in thin films of strongly disordered superconductors), where  finite value of   $\lambda_{2D}$ does not play any role.

Phenomenological approach to the problem of vortices moving in the film, similar to one used in Ref. \cite{ES}, was developed in the paper~\cite{Nelson1995}  (see also earlier paper~\cite{LarkinKhmelnitsky}).  
Here we  develop alternative approach based on the paper by M{\"u}ller and Ioffe~\cite{IM2004}
(see also  papers~\cite{MullerPankov,Pankov}), where the problem of Coulomb gap  was studied using spin glass theory methods, and a phase transition to the non-ergodic state with broken replica symmetry was predicted. However, unlike the paper~\cite{IM2004}, we will not assume that the theory can be described by a purely \textit{local} matrix model neglecting the spatial fluctuations of matrix fields describing the glass phase.

\textit{2. The model and mean field theory.} 
We will use  model assumption that vortices can occupy positions of a discrete regular lattice with lattice constant $a$.
The configuration of  vortices will be described by <<occupation numbers>> of each cite $\{n_{\boldsymbol{r}}\}$.
External magnetic  field $B$ leads to a finite vortex density $\left\langle n_{\boldsymbol{r}}\right\rangle \equiv K=Ba^{2}/\Phi_{0}$. We  neglect  anti-vortices as well as vortices with charge $n_{\boldsymbol{r}}>1$. Finite concentration of vortices will be fixed by the chemical potential $\mu$. Finally,  disorder in our model will be described by  random energy of a vortex core $u_{\boldsymbol{r}}$ those correlation function is $\overline{ u_{\boldsymbol{r}}u_{\boldsymbol{r}^{\prime}}} =W^{2}\delta_{\boldsymbol{r}\boldsymbol{r}^{\prime}}$. 
It leads to the following Hamiltonian:
\begin{equation}
H=\frac{1}{2}\sum_{\boldsymbol{r},\boldsymbol{r}^{\prime}}\delta n_{\boldsymbol{r}}J_{\boldsymbol{r}\boldsymbol{r}^{\prime}}\delta n_{\boldsymbol{r}^{\prime}}+\sum_{\boldsymbol{r}}(u_{\boldsymbol{r}}-\mu)\delta n_{\boldsymbol{r}},
\end{equation}
where $\delta n_{\boldsymbol{r}}\equiv n_{\boldsymbol{r}}-K$ and $J_{\boldsymbol{r}\boldsymbol{r}^{\prime}}=U_{0}\ln\frac{L}{|\boldsymbol{r}-\boldsymbol{r}^{\prime}|}$. Disorder strength is assumed to be large, $W \gg U_0$. In fact, in the superconductors we consider, $W \sim U_0$; in the conclusion we will discuss why the model assumption $W \gg U_0$ will not affect our main results.
We average the free energy over the disorder utilizing the replica trick, and perform a Hubbard-Stratonovich transformation of  a non-local term introducing the auxiliary field  $\varphi$ (it has  the meaning of the dual variable to the superconducting phase). In result, we arrive at the following expression for the partition function:
\begin{multline}
\overline{Z^n} =\int{\cal D}\varphi\exp\left(-\frac{1}{2}\varphi(\beta\hat{J})^{-1}\varphi\right)\times\\
\times\prod_{\boldsymbol{r}}\Tr_{\text{v}}\exp\left(\sum_{a}(\beta\mu+i\varphi_{\boldsymbol{r}}^{a})\delta n_{\boldsymbol{r}}^{a}+\frac{\beta^{2}W^{2}}{2}\delta n_{\boldsymbol{r}}\hat{{\cal I}}\delta n_{\boldsymbol{r}}\right),
\label{eq:PartitionFunction}
\end{multline}
where we denoted $\Tr_{\text{v}}\equiv\sum_{n_{\boldsymbol{r}}=0,1}$; latin indices numerate replicas $a = 1, \dots, n$ ($n \to 0$), matrix ${\cal I}^{ab}=1$ describes the quenched disorder equivalent for all replicas, and the interaction $\hat{J}=\delta^{ab}J_{\boldsymbol{r}\boldsymbol{r}^{\prime}}$ is diagonal in replica space. 
It is worth noting that in this expression, the <<vortex>> part of the action appears now purely local.

We characterize the glass state by means of diagonal in coordinate space (yet coordinate-dependent) matrix ${\cal G}_{\boldsymbol{r}}^{ab}$, which describes the correlations of slowly varying in space part of  bilinear combination of fields
$=-\varphi_{\boldsymbol{r}}^{a}\varphi_{\boldsymbol{r}}^{b}$. Glass transition corresponds to  spontaneous replica symmetry breaking in such a matrix. The order parameter is introduced utilizing the following identity (the integral over ${\cal Q}$ is taken along imaginary axis):
\begin{multline}
1=\int{\cal D}{\cal G}\prod_{\boldsymbol{r}}\delta({\cal G}_{\boldsymbol{r}}^{ab}+\varphi_{\boldsymbol{r}}^{a}\varphi_{\boldsymbol{r}}^{b})=\\
=\int{\cal D}{\cal G}{\cal D}{\cal Q}\exp\left(-\frac{1}{2}\Tr\left(\hat{{\cal G}}\hat{{\cal Q}}\right)-\frac{1}{2}\varphi\hat{{\cal Q}}\varphi\right)
\end{multline}
The fluctuations of $\varphi$ field are described by the propagator with the screening length $l\sim a\sqrt{W/U_{0}}$. On the other hand, it is reasonable to assume that the fluctuations of the order parameter $\hat{{\cal G}}_{\boldsymbol{r}}$ will be correlated on much larger spatial scales in the glass phase and in the vicinity of the transition.

In order to deal with the interaction between vortex occupation numbers $n_{\mathbf{r}}^{a}$ and 
$\varphi_{\boldsymbol{r}}^{a}$ field, we expand $\exp(i\sum_{a}\varphi_{\boldsymbol{r}}^{a}\delta n_{\boldsymbol{r}}^{a})$ in the Taylor series and rewrite arbitrary term in the momentum representation:
\begin{equation}
e^{i\sum_{a}\varphi_{\boldsymbol{r}}^{a}\delta n_{\boldsymbol{r}}^{a}}=\sum_{k=0}^{\infty}\sum_{\boldsymbol{q}_{1}+\dots+\boldsymbol{q}_{k}=0}\frac{i^{k}}{k!}\delta n_{\boldsymbol{r}}^{a_{1}}\dots\delta n_{\boldsymbol{r}}^{a_{k}}\varphi_{\boldsymbol{q}_{1}}\dots\varphi_{\boldsymbol{q}_{k}}
\label{eq:expansion}
\end{equation}
We wish to describe fluctuations of soft modes of the order parameter $\hat{{\cal G}}_{\boldsymbol{r}}$ with the wavevectors much smaller than  typical wavevectors of $\varphi$ fields, that is $q_{i}\sim l^{-1}$.
The main contribution to such fluctuations  come from the terms in \eqref{eq:expansion}, where some pairs of wavevectors are anomalously close $|\mathbf{q}_{i}+\mathbf{q}_{j}| \ll l^{-1}$ --- such <<contractions>> will then be replaced by ${\cal G}_{\boldsymbol{q}_{i}+\boldsymbol{q}_{j}}^{a_{i}a_{j}}$ (meaning that the total wavevector is small).
So, in order to obtain the leading contribution to  fluctuations of the slow modes, we need to consider all <<contractions>> of $\varphi$ fields  in this expression. 
The terms with odd $k$ then describe the interaction between the slow and fast modes, and can be neglected in the leading order. This allows us to replace the interaction between the field
$\varphi(\mathbf{r})$ and vortex degrees of freedom by the local interaction between vortices described by the term $\delta n_{\boldsymbol{r}}\hat{{\cal G}}_{\boldsymbol{r}}\delta n_{\boldsymbol{r}}/2$ in the exponent.

We finally do the remaining Gaussian integral over $\varphi(\mathbf{r})$, and arrive at the following field theory describing fluctuations of slow modes of the matrix order parameter:
\begin{equation}
\overline{Z^{n}} =\int{\cal D}{\cal G}{\cal D}{\cal Q}\exp\left(-nS[\hat{{\cal G}},\hat{{\cal Q}}]\right)
\end{equation}
\begin{equation}
nS[\hat{{\cal G}},\hat{{\cal Q}}]=\frac{1}{2}\Tr(\hat{{\cal G}}\hat{{\cal Q}})+\frac{1}{2}\Tr\ln(1+\beta\hat{J}\hat{{\cal Q}})+\beta n\sum_{\boldsymbol{r}}F_{\text{v}}[\hat{{\cal G}}_{\boldsymbol{r}}]
\label{eq:Action}
\end{equation}
where the local part of the free energy is given by the expression:
\begin{equation}
e^{-\beta nF_{\text{v}}[\hat{{\cal G}}]}=\Tr_{\text{v}}\exp\left(\frac{1}{2}\delta n(\beta^{2}W^{2}\hat{{\cal I}}+\hat{{\cal G}})\delta n+\beta\mu\sum_{a}\delta n^{a}\right)
\label{eq:LocalHamiltonian}
\end{equation}

We begin the analysis of the action \eqref{eq:Action} by studying the spatially homogeneous saddle points:
\begin{equation}
\frac{\delta S}{\delta\hat{{\cal G}}}=\frac{1}{2}(\hat{\cal Q}-\hat{Q})=0,\quad Q_{ab}=\left\langle \delta n_{a}\delta n_{b}\right\rangle_{\cal{G}},
\label{eq:SaddlePointQ}
\end{equation}
where $\hat{Q}$ is the  density correlation function calculated in the local model \eqref{eq:LocalHamiltonian}. 

The second saddle-point equation acquires the following form, in agreement with the definition of the ${\cal G}$ matrix:
\begin{equation}
\frac{\delta S}{\delta\hat{\cal Q}}=\frac{1}{2}(\hat{{\cal G}}+\hat{G}_{\boldsymbol{r}\boldsymbol{r}})=0,\quad \hat{G}=((\beta\hat{J})^{-1}+\hat{\cal Q})^{-1}
\label{eq:SaddlePointG}
\end{equation}

To illustrate  the role of the $\hat{G}$ matrix, let us introduce into a system a pair of infinitesimal vortices with charges $q_{1,2} \ll 1$ to the points $\boldsymbol{r}_{1,2}$ in the replicas $a_{1,2}$. It corresponds to the following perturbation of the system Hamiltonian:
\begin{equation}
V=\sum_{\boldsymbol{r}}\left(q_{1}J_{\boldsymbol{r}_{1}\boldsymbol{r}}\delta n_{\boldsymbol{r}}^{a_{1}}+q_{2}J_{\boldsymbol{r}_{2}\boldsymbol{r}}\delta n_{\boldsymbol{r}}^{a_{2}}\right)+q_{1}q_{2}J_{\boldsymbol{r}_{1}\boldsymbol{r}_{2}}
\end{equation}
 Free energy response to such a perturbation  determines  interaction energy between  two added vortices.
After the Hubbard-Stratanovich transformation, one finds a correction to the action in  the exponent in Eq. \eqref{eq:PartitionFunction}, equal to $-i(q_{1}\varphi_{\boldsymbol{r}_{1}}^{a}+q_{2}\varphi_{\boldsymbol{r}_{2}}^{b})$. 
Doing then the Gaussian integral over $\varphi$, one finds the following additional contribution to the expression \eqref{eq:Action}:
\begin{equation}
\delta S=\frac{1}{2}\left(q_{1}^{2}G_{\boldsymbol{r}_{1}\boldsymbol{r}_{1}}^{a_1a_1}+q_{2}^{2}G_{\boldsymbol{r}_{2}\boldsymbol{r}_{2}}^{a_2a_2}\right)+q_{1}q_{2}G_{\boldsymbol{r}_{1}\boldsymbol{r}_{2}}^{a_1a_2}
\end{equation}
It means that the average value of  $\hat{G}$ matrix  can be identified with the effective  interaction between two  <<infinitesimal>> vortices:
\begin{equation}
U_{a_{1}a_{2}}^{\text{(eff)}}(\boldsymbol{r}_{1},\boldsymbol{r}_{2})\equiv\left.\frac{\partial^{2}F}{\partial q_{1}\partial q_{2}}\right|_{q_{1,2}=0}=T\left\langle G_{\boldsymbol{r}_{1}\boldsymbol{r}_{2}}^{a_1a_2}\right\rangle 
\label{eq:VortexInteraction}
\end{equation}

Finally, we  write the equation for the chemical potential:
\begin{equation}
\left\langle \frac{\partial S}{\partial(\beta\mu)}\right\rangle =\sum_{a}\left\langle \delta n^{a}\right\rangle _{{\cal G}}=0
\end{equation}
As $W$ is assumed to be the largest parameter is the problem,  chemical potential in the leading order is determined by the <<bare>> density of states $\nu(u)=\exp(-u^{2}/2W^{2})/\sqrt{2\pi}W$ via the following equation:
\begin{equation}
1-2K=\int\nu(u)du\tanh\frac{\beta(u-\mu)}{2}\approx\int\nu(u)du\cdot{\rm sign}(u-\mu)
\end{equation}
which yields  asymptotic expressions:
\begin{equation}
\mu\approx -W\cdot\begin{cases}
\sqrt{2\pi}\left(\frac{1}{2}-K\right), & |K-1/2|\ll1\\
\left(2\ln\frac{1}{\sqrt{2\pi}K}\right)^{1/2}, & K\ll1
\end{cases}
\end{equation}

\textit{3. High-temperature phase and glass transition.}
We begin  from the high-temperature phase corresponding to the replica-symmetric solutions ${\cal G}_{ab}={\cal G}_{0}\delta_{ab}+{\cal G}_{1}{\cal I}_{ab}$ (and the same for ${\cal Q}$). 
Since the vortex variables $\delta n$ are similar to the Ising spin variables, the following identity can be written: $\delta n^{2}=\delta n(1-2K)+K(1-K)$ (at $K=1/2$ it reads $s^{2}=1/4$ for Ising spin variables). As a consequence, the diagonal part ${\cal G}_{0}$ simply renormalizes the chemical potential $\mu \mapsto \mu+T{\cal G}_{0}\left(\frac{1}{2}-K\right)$, while  off-diagonal part renormalizes the disorder strength $W\mapsto\sqrt{W^{2}+T^{2}{\cal G}_{1}}$. Both effects are actually negligible because $\mu\sim W \gg T,U_0$.

We obtain the following solutions:
\begin{equation}
{\cal Q}_{0}=\int\frac{\nu(u)du}{\left(2\cosh\frac{\beta(u-\mu)}{2}\right)^{2}}\approx T\nu_0.
\end{equation}
\begin{equation}
{\cal Q}_{1}=K(1-K)-{\cal Q}_{0}.
\end{equation}
Due to  large value of  $W$, the density of states $\nu(u)$ can actually be replaced with a constant:
\begin{equation}
\nu_{0}\equiv\nu(\mu)\approx\begin{cases}
\frac{1}{\sqrt{2\pi}W}, & |K-1/2|\ll1\\
\frac{K}{W}\left(2\ln\frac{1}{\sqrt{2\pi}K}\right)^{1/2}, & K\ll1
\end{cases}
\end{equation}

As a result, the screening appears in the propagator  $G_{ab}(\boldsymbol{k})=G_{0}(\boldsymbol{k})\delta_{ab}+G_{1}(\boldsymbol{k}){\cal I}_{ab}$:
\begin{equation}
G_{0}(\boldsymbol{k}) = \frac{2\pi\beta U_{0}}{k^{2}+l^{-2}},\quad G_{1}(\boldsymbol{k})=-{\cal Q}_{1} G_{0}^{2}(\boldsymbol{k}) / a^{2}
\end{equation}
with $l=a(2\pi\nu_{0}U_{0})^{-1/2}\sim a\sqrt{W/U_{0}}$. Finally, the order parameter is:
\begin{equation}
{\cal G}_{0}\approx-\frac{\beta U_{0}}{2}\ln\frac{1}{\nu_{0}U_{0}},\quad{\cal G}_{1}\approx\frac{\beta^{2}U_{0}}{\nu_{0}}K(1-K)
\end{equation}

In order to study the stability of the replica-symmetric solution and deduce the freezing transition temperature, one needs to study the Hessian --- the quadratic expansion of the action \eqref{eq:Action}:
\begin{multline}
nS^{(2)}[\delta\hat{{\cal G}},\delta\hat{{\cal Q}}]=\frac{1}{2}\Tr(\delta\hat{{\cal G}}\delta\hat{{\cal Q}})-\frac{1}{4}\Tr(\hat{G}\delta\hat{{\cal Q}}\hat{G}\delta\hat{{\cal Q}})\\-\frac{1}{8}\sum_{\boldsymbol{r}}Q_{(a_1 b_1)(a_2 b_2)}\delta{\cal G}_{\boldsymbol{r}}^{a_1 b_1}\delta{\cal G}_{\boldsymbol{r}}^{a_2 b_2}
\end{multline}
where we have introduced the following correlation function:
\begin{multline}
Q_{(a_1 b_1)(a_2 b_2)}\equiv\left\langle \delta n_{a_1}\delta n_{b_1}\delta n_{a_2}\delta n_{b_2}\right\rangle _{{\cal G}}\\
-\left\langle \delta n_{a_1}\delta n_{b_1}\right\rangle _{{\cal G}}\left\langle \delta n_{a_2}\delta n_{b_2}\right\rangle _{{\cal G}}
\end{multline}

Upon lowering the temperature, a singularity appears in the replicon mode. This mode corresponds to the linear subspace of matrices subject to the following two constraints:  $\delta{\cal G}_{aa} = 0$ and $\sum_{a}\delta{\cal G}_{ab}=0$. The action for the replicon fluctuations then reads:
\begin{equation}
nS^{(2)}\approx\int\frac{(d\boldsymbol{q})}{4a^{2}}{\rm tr}\begin{pmatrix}\delta\hat{{\cal G}}_{\boldsymbol{q}} & \delta\hat{{\cal Q}}_{\boldsymbol{q}}\end{pmatrix}\begin{pmatrix}-Q_{22} & 1\\
1 & -{\cal B}_2(\boldsymbol{q})
\end{pmatrix}\begin{pmatrix}\delta\hat{{\cal G}}_{-\boldsymbol{q}}\\
\delta\hat{{\cal Q}}_{-\boldsymbol{q}}
\end{pmatrix},
\label{eq:ActionQuadratic}
\end{equation}
where symbol ${\rm tr}$ corresponds to the trace  w.r.t. replica space only, and the following notations were introduced:
\begin{equation}
{\cal B}_2(\boldsymbol{q})=\int(d\boldsymbol{k})G_{0}(\boldsymbol{k})G_{0}(\boldsymbol{k}+\boldsymbol{q})\approx \pi(\beta U_{0}l)^{2}\left(1-q^{2}l^{2}/6\right)
\label{eq:RepliconBEigenvalue}
\end{equation}

\begin{equation}
Q_{22}=\int\frac{\nu(u)du}{\left(2\cosh\frac{\beta(u-\mu)}{2}\right)^{4}}\approx T\nu_{0}/6
\label{eq:RepliconQEigenvalue}
\end{equation}

Quadratic expansion \eqref{eq:ActionQuadratic} corresponds to the ladder summation of diagram series for a four-point Green function of the $\varphi$ field. 
The action \eqref{eq:ActionQuadratic} yields the following propagators:
\begin{eqnarray}
\label{G}
\left\langle\left\langle {\cal G}_{\boldsymbol{r}}^{ab}{\cal G}_{\boldsymbol{r}^{\prime}}^{a^{\prime}b^{\prime}}\right\rangle\right\rangle _{\boldsymbol{q}}\approx& \frac{12 \beta a^{2}/\nu_{0}}{\tau+q^{2}l^{2}/6}\,
\mathbb{P}_{bb^{\prime}}^{aa^{\prime}}, \\
\left\langle\left\langle {\cal Q}_{\boldsymbol{r}}^{ab}{\cal Q}_{\boldsymbol{r}^{\prime}}^{a^{\prime}b^{\prime}}\right\rangle\right\rangle _{\boldsymbol{q}}\approx& \frac{a^{2}\nu_{0}T^{2}/3T_{c}}{\tau+q^{2}l^{2}/6}\,
\mathbb{P}_{bb^{\prime}}^{aa^{\prime}},
\label{eq:RepliconPropagators}
\end{eqnarray}
where we have introduced the freezing temperature $T_c \equiv U_0/12$; at this temperature the value $\tau\equiv T/T_{c}-1$ changes sign, and the instability appears in the theory \eqref{eq:ActionQuadratic} leading to the spontaneous symmetry breaking. The tensor $\mathbb{P}_{bb^{\prime}}^{aa^{\prime}}$ is the projector on to the replicon mode.

As it is shown in the Supplementary 1, the correlation function $\left\langle \left\langle {\cal Q}{\cal Q}\right\rangle \right\rangle $ has the following physical meaning: it describes the long-wavelength asymptotic of the mean square fluctuation of the polarizability:
\begin{equation}
\overline{\left\langle \delta n_{\boldsymbol{r}}\delta n_{\boldsymbol{r}^{\prime}}\right\rangle ^{2}}=\lim_{n\to0}\frac{1}{n(n-1)}\sum_{a\neq b}\left\langle \left\langle \hat{{\cal Q}}_{\boldsymbol{r}}^{ab}\hat{{\cal Q}}_{\boldsymbol{r}^{\prime}}^{ab}\right\rangle \right\rangle 
\end{equation}
Finally, replica structure of the projector onto the replicon mode gives  additional factor of $\lim_{n\to0}\mathbb{P}_{bb}^{aa} / n(n-1) =3/2$ to the expression \eqref{eq:RepliconPropagators}.

In the vicinity of the transition, when $\tau \ll 1$,  quadratic part of the action \eqref{eq:ActionQuadratic} can be approximately diagonalized by the following transformation:
\begin{equation}
\begin{pmatrix}\hat{\Psi}\\
\hat{\Phi}
\end{pmatrix}=\begin{pmatrix}1/2 & 1/2Q_{22}\\
1/2 & -1/2Q_{22}
\end{pmatrix}\begin{pmatrix}\delta\hat{{\cal G}}\\
\delta\hat{{\cal Q}}
\end{pmatrix},
\end{equation}
The mode $\Psi$ appears to be soft, and the mode $\Phi$ is gapped and thus can be neglected.
Expanding the functional w.r.t.  $\hat{\Psi}$, we arrive at (the details of the calculation are given in the Supplementary 2) the following Ginzburg-Landau functional:
\begin{multline}
nS[\hat{\Psi}]=\nu_{0}T_{c}\Bigg(\frac{1}{24}\Tr(\tau\hat{\Psi}^{2}+(\nabla\hat{\Psi})^{2}l^{2}/6) - \\
-\frac{1}{2160}\left(7\Tr\hat{\Psi}^{3}+6\sum_{ab,\boldsymbol{r}}\Psi_{ab,\boldsymbol{r}}^{3}\right)-\frac{1}{2016}\sum_{ab,\boldsymbol{r}}\Psi_{ab,\boldsymbol{r}}^{4}\Bigg)
\label{Smulti}
\end{multline}
Despite the large screening length $l \gg a$ in our problem, all the coefficients in front of the non-linear terms are of the same order $\sim \nu_0 T_c \sim U_0 / W$. 
As a consequence, the derived Ginzburg-Landau theory lacks a small parameter, and the Ginzburg region where the fluctuation effects are strong  is of the width $\text{Gi}=O(1)$; thus  the mean field theory is inapplicable in the vicinity of the transition. 
The same conclusion applies to the three-dimensional counterpart of the same problem which was studied in  Ref.~\cite{IM2004}.  Strong critical fluctuations prevent us from the study of the critical region itself, therefore we switch to the low-temperature phase of the model, where  fluctuation effects are suppressed by small factor
$T/T_c \ll 1 $.

\textit{4. Low-temperature phase in the 1-step replica symmetry breaking approximation.}
The ratio between coefficients in front of two cubic terms in the action \eqref{Smulti}, $c_1 / c_2 = 6/7 < 1$, which
suggests that the full continuous replica symmetry breaking scheme due to Parisi~\cite{MezardBook} should be used; 
if the same ratio
would be $> 1$, then 1-step replica symmetry breaking scheme (1-RSB)~\cite{GrossKanterSompolinsky} would be sufficient.
In our problem the ratio $c_1 / c_2$ is quite close to unity, thus we will try to apply the 1-RSB approximation and show
\textit{a posteriori} that the obtained solution is a very good one \textit{numerically}. 1-RSB scheme suggests the following form for the matrices:
\begin{equation}
{\cal G}_{ab}={\cal G}_{0}\delta_{ab}+{\cal G}_{1}{\cal R}_{ab}+{\cal G}_{2}{\cal I}_{ab}
\end{equation}
\begin{equation}
{\cal Q}_{ab}={\cal Q}_{0}\delta_{ab}+\frac{1}{m}({\cal Q}_{1}-{\cal Q}_{0}){\cal R}_{ab}+{\cal Q}_{2}{\cal I}_{ab}
\end{equation}
The auxiliary matrix ${\cal R}_{ab}=\delta_{\left[a/m\right],\left[b/m\right]}$ (here $[\dots]$ denotes the integer part) is a block-diagonal matrices with diagonal blocks of size $m \times m$ being filled with ones, while off-diagonal blocks are filled with zeroes. In the replica limit $n \to 0$, the parameter $m \in (0,1)$ becomes an additional variational parameter of our theory.
The Green function $\hat{G}$, see Eq.\eqref{eq:SaddlePointG}, is parametrized in the same fashion:
\begin{equation}
G_{ab}(\boldsymbol{k})=G_{0}(\boldsymbol{k})\delta_{ab}+\frac{1}{m}(G_{1}(\boldsymbol{k})-G_{0}(\boldsymbol{k})){\cal R}_{ab}+G_{2}(\boldsymbol{k}){\cal I}_{ab},
\end{equation}
with
\begin{equation}
G_{0,1}(\boldsymbol{k})=\frac{2\pi\beta U_{0}}{k^{2}+l_{0,1}^{-2}},\quad G_{2}(\boldsymbol{k})=-{\cal Q}_{2}G_{1}^{2}(\boldsymbol{k})/a^{2},
\label{eq:LowTGreenFunction}
\end{equation}
where two different screening lengths appears $l_{0,1}=a(2\pi\beta U_{0}{\cal Q}_{0,1})^{-1/2}$. 
The first group of saddle point equations, Eq. \eqref{eq:SaddlePointG}, reads:
\begin{equation}
\begin{cases}
{\cal G}_{0} & \approx-\beta U_{0}\ln(l_{0}/a)\approx\beta U_{0}\ln(\beta U_{0}{\cal Q}_{0})/2\\
{\cal G}_{1} & =\beta U_{0}\ln(l_{0}/l_{1})/m=\beta U_{0}\ln({\cal Q}_{1}/{\cal Q}_{0})/2m\\
{\cal G}_{2} & =\pi{\cal Q}_{2}(\beta U_{0}l_{1}/a)^{2}=\beta U_{0}{\cal Q}_{2}/2{\cal Q}_{1}
\end{cases}
\label{eq:1RSBEquationsG}
\end{equation}
The second group, Eq. \eqref{eq:SaddlePointQ}, in the limit of $W \gg U_0$, can be expressed via the auxiliary function $f_{\text{v}}(m,{\cal G}_{1})$ (see Supplementary 3 for details):
\begin{equation}
\begin{cases}
{\cal Q}_{0} & =\frac{\nu_{0}T}{1-m}\partial f_{\text{v}}/\partial{\cal G}_{1}\\
{\cal Q}_{1} & =\nu_{0}T\\
{\cal Q}_{2} & =K(1-K)+\left(\frac{1}{m}-1\right){\cal Q}_{0}-\frac{1}{m}{\cal Q}_{1}
\end{cases},
\label{eq:1RSBEquationsQ}
\end{equation}
and the auxiliary function reads:
\begin{multline}
f_{\text{v}}(m,{\cal G}_{1})=\frac{2}{m}\int dz\Big(\ln\Xi(z,m,{\cal G}_{1})\\
-m \ln 2 \cosh\frac{z}{2}-\frac{m^{2}{\cal G}_{1}}{8}\Big)
\label{eq:fv}
\end{multline}
\begin{equation}
\Xi(z,m,{\cal G}_{1})=\int\frac{dye^{-y^{2}/2{\cal G}_{1}}}{\sqrt{2\pi{\cal G}_{1}}}\left[2\cosh\frac{y-z}{2}\right]^{m}
\end{equation}
Last equation of the group \eqref{eq:1RSBEquationsQ} is  trivial consequence of the fact that diagonal elements are fixed via the relation ${\cal Q}_{aa}=K(1-K)$; the second equation suggests that the screening length $l_1$ coincides with the screening length in the replica-symmetric phase.
Finally, to close the whole system of equations  we need to add stationary equation for the 1-RSB parameter $m$, which can be written in the following form:
\begin{equation}
-\frac{6 T_c}{m T}\left(1-\frac{1}{1-m}\frac{\partial f_{\text{v}}}{\partial{\cal G}_{1}}\right)+{\cal G}_{1}\frac{\partial f_{\text{v}}}{\partial{\cal G}_{1}}-m\frac{\partial f_{\text{v}}}{\partial m}=0
\label{eq:1RSBEquationM}
\end{equation}
Among seven equations  (\ref{eq:1RSBEquationsG}, \ref{eq:1RSBEquationsQ}, \ref{eq:1RSBEquationM}),  only equations for $(m,{\cal G}_{1},{\cal Q}_{0})$ are nontrivial.

At low temperatures the system of equations (\ref{eq:1RSBEquationsG}, \ref{eq:1RSBEquationsQ}, \ref{eq:1RSBEquationM}) obeys the following solution (see Supplemental 3.1 for details):
\begin{equation}
m\approx1.09\,(T/T_{c}),\quad{\cal G}_{1}\approx61.0\,(T_{c}/T)^{2}
\label{eq:mLowT}
\end{equation}
\begin{equation}
{\cal Q}_{0}\approx1.43\cdot10^{-5}\,\nu_{0}T
\label{eq:Q0LowT}
\end{equation}

\textit{5. Physical properties of the low-temperature phase.}
As we have shown above (Eq. \eqref{eq:VortexInteraction}), the $\hat{G}$ matrix describes the interaction energy for two probe vertices introduced to the system.
It is known from the theory of spin glasses~\cite{MezardBook}  that  replica symmetry breaking physically corresponds to the breaking of the ergodicity and dependence of the system state on its history. 
In particular,  two protocols are commonly considered: the  <<Zero Field Cooling>> (ZFC), which corresponds to introducing the probe vertices \textit{after} freezing into the glass state, and <<Field Cooling>>, which corresponds to introducing the vertices \textit{before} the freezing. In the replica technique it corresponds to the following two response functions:
\begin{multline}
U_{\text{ZFC}}^{\text{(eff)}}(\boldsymbol{r}_{1},\boldsymbol{r}_{2})=\lim_{b\to a}\left[U_{aa}^{\text{(eff)}}(\boldsymbol{r}_{1},\boldsymbol{r}_{2})-U_{ab}^{\text{(eff)}}(\boldsymbol{r}_{1},\boldsymbol{r}_{2})\right]\\
=T G_{0}(\boldsymbol{r}_{1}-\boldsymbol{r}_{2})
\end{multline}
\begin{equation}	
U_{\text{FC}}^{\text{(eff)}}(\boldsymbol{r}_{1},\boldsymbol{r}_{2})=\frac{1}{n}\sum_{ab}U_{ab}^{\text{(eff)}}(\boldsymbol{r}_{1},\boldsymbol{r}_{2})=T G_{1}(\boldsymbol{r}_{1}-\boldsymbol{r}_{2})
\end{equation}
The extreme smallness of $\mathcal{Q}_0$ (Eq. \eqref{eq:Q0LowT}) and the relation~\eqref{eq:LowTGreenFunction} 
leads to  very large value of a screening length for the ZFC-response in the glass phase $l_0 \approx 260 l_1$ ($l_1$ coincides with the screening length in the high-temperature phase). In any experimentally feasible situation, such value $l_0$ can be considered as infinity.
As a result, at low temperatures $T \ll T_c$  logarithmic interaction between vortices is restored, and such phase is characterized by nonzero superfluid stiffness:
\begin{equation}
\rho_{\text{ZFC}}^{\text{(s)}}=\frac{T}{4\pi^2}\lim_{k\to0}k^{2} G_{0}(\boldsymbol{k})\simeq \frac{U_{0}}{2 \pi}
\label{eq:SuperfluidStiffness}
\end{equation}

Another important physical quantity in the problem is the distribution function of the local potential of an individual vortex, $P(u)$. 
Detailed calculation of this distribution function at low temperatures is described in Supplemental 2.2; here we present the approximate result, which is valid at low temperatures:
\begin{equation}
P\left(h\equiv\frac{u-\widetilde{\mu}}{T_{c}}\right)\approx\nu_{0}\cdot\frac{1}{2}\text{erfc}\left(3.03 - 0.09 |h|\right).
\label{eq:Ph}
\end{equation} 

At low temperatures, the gap develops in the distribution function. The half-width of the gap is of the order of $\sim 30 T_c = 2.5 U_0$, and the absolute value of the density of states exactly at the chemical potential is negligibly small $\sim 10^{-5}$, albeit non-zero. This small value is the actual reason behind the small value of ${\cal Q}_{0}$ (see Eq. \eqref{eq:Q0LowT}), which can be expressed as follows:
\begin{equation}
{\cal Q}_{0}=\int\frac{P(u)du}{\left(2\cosh\frac{\beta(u-\widetilde{\mu})}{2}\right)^{2}} \approx T P_0,\quad P_0 \equiv P(\widetilde{\mu}).
\end{equation}
The last equality takes into account the fact that the density of states is nearly constant at the scales $|u - \widetilde{\mu}| \sim T$.

The low-temperature phase in the 1-RSB approximation is unstable --- the replicon mode, which is responsible for additional  replica symmetry breaking, has a negative eigenvalue. 
However, Eqs. (\ref{eq:ActionQuadratic}-\ref{eq:RepliconPropagators}) are still applicable to the replicon mode, with only difference being that the screening length $l$ should be replaced by $l_0$ and the density of states $\nu_0$ should be replaced by its renormalized value $P_0$. This is possible because the density of states, despite having a large gap of width $\sim 30 T_c$, can be considered almost constant at the scales $\sim T$ that we are interested in.  In particular, at $T \ll T_c$ the value $\tau = T/T_c - 1 \approx -1$, and thus the mode $q = 0$ is indeed unstable. However, due to the value $l_0 = a / \sqrt{2 \pi \beta U_0 {\cal Q}_0}$ having a large numerical factor $\sim 250$, the phase volume of unstable modes $q \lesssim 1 / l_0$ appears to be extremely small. It leads to the natural assumption that approximations we have made here can be used to describe the system with a good precision.

Entropy in the 1-RSB approximation can be written as 
\begin{equation}
S=\nu_{0}T\left[f_{\text{v}}(m,{\cal G}_{1})+\frac{1}{2}m\frac{\partial f_{\text{v}}}{\partial m}-{\cal G}_{1}\frac{\partial f_{\text{v}}}{\partial{\cal G}_{1}}+\frac{\pi^{2}}{3}\right]-3\beta T_{c}{\cal Q}_{0}
\end{equation}
At low temperatures, the behavior of the entropy is discussed in Supplemental 2.3; here we briefly state the results. At zero temperature the entropy is negative, but its absolute value is extremely small (which also stems from the low density of states):
\begin{equation}
S(T=0)=-3\beta T_{c}{\cal Q}_{0}\approx-4.29\cdot10^{-5}\nu_{0}T_{c}
\end{equation}

The freezing transition of the vortex glass can also be considered from another point of view, as a statistical mechanics problem of a particle in the logarithmically correlated random potential \cite{LeDoussal2001}. Indeed, albeit the <<bare>> disorder is short-range correlated, the \textit{effective} random potential probed by a separate vortex has the following form:
\begin{equation}
u_{\text{eff}}(\boldsymbol{r})=u(\boldsymbol{r})+\sum_{\boldsymbol{r}^{\prime}}J_{\boldsymbol{r}\boldsymbol{r}^{\prime}}\delta n_{\boldsymbol{r}^{\prime}},
\end{equation}
and its fluctuations on the scales $l_{0}\gg r \gg  l_{1}$ could be estimated (utilizing \eqref{eq:mLowT} and
\eqref{eq:LowTGreenFunction}) as follows:
\begin{multline}
\overline{\left\langle u_{\text{eff}}(\boldsymbol{r})-u_{\text{eff}}(0)\right\rangle ^{2}}\underset{|\boldsymbol{r}-\boldsymbol{r}^{\prime}|\gg l_1}{=}\frac{2 (1-m)}{m}T^{2}(G_{0}(\boldsymbol{r})-G_{0}(0)) \\
\approx 11T_{c}^{2}\ln\frac{r}{a},\quad T \ll T_c
\end{multline}
The freezing criterion \cite{LeDoussal2001} suggests that system should be in the frozen state at $T < \sqrt{5.5 T_c^2} \propto T_c$, which confirms our conclusion that glass state is realized at $T \leq T_c$.

\textit{6. Conclusions.}
The results of our work can be (qualitatively) applied to the analysis of~\cite{Yazdani2013,Sacepe2019} where the experiments were performed with  highly disordered superconducting films of InO$_x$ and MoGe. In the paper~\cite{Yazdani2013}  kinetic inductance $L_K$ at low frequencies was studied in the broad range of magnetic field and temperature, and it was found that the criterion for the superconducting state with non-zero superfluid stiffness $\rho_s \propto 1/L_K$ is close to the criterion for the Berezinskii-Kosterlitz-Thouless transition, i.e.  $\rho_s/T_c \approx \mathrm{const}$.
In Ref.~\cite{Sacepe2019}  transport critical current $j_c$ was studied for films of InO$_x$ at $T \ll T_c$ and magnetic fields close to the upper critical field $H_{c2}$. It was found that the dependence $j_c(B)$ is very close to the ``mean-field'' result $j_c(B) \propto (H_{c2}-B)^{3/2}$. The semi-qualitative arguments explaining this behaviour were presented in the paper~\cite{Sacepe2019}. 
However, the main conclusion that can be drown from experiments~\cite{Yazdani2013,Sacepe2019} is the very existence of a non-zero superfluid stiffness in systems subject to strong magnetic field.  It can only be realised if the dense system of vortices possesses some sort of glass phase.  
In the present Letter the first analytic approach to describe such a vortex glass state was developed.

We have utilized the model assumption of a very strong local disorder, $W \gg U_0$. In fact, in such systems and in magnetic fields $B \ll H_{c2}$, the disorder strength is of the order of $W \sim 0.5 U_0$. However, dropping the assumption $W \gg U_0$ won't affect our main results for the glass phase. It is due to the large width (see Eq. \eqref{eq:Ph}) of the ``gap'' in the local pinning energies distribution function $P(h)$, which has the half-width $30 T_c \approx 2.5 U_0$. Moreover, as magnetic field $B$ approaches $H_{c2}$, the pinning energy of individual vortices drops as $1 - B / H_2$, while the interaction strength drops as $(1 - B / H_{c2})^2$, which further increases the ratio $W / U_0$.

Formally, the obtained 1-RSB solution is unstable, which signals that the theory utilizing the continuous Parisi scheme should be developed. However, the difference between such a full theory and the one presented in this work is expected to be extremely small, which is suggested by the value of the entropy per cite $ - S_0 \approx 10^{-5}$. Furthermore, these discrepancies should in fact be described using the dynamical spin glass theory, since fluctuations at the scales of the order of $l_0$ at $T \ll T_c$ cannot occur in a thermodynamically equilibrium fashion. 
Finally, we wish to point that in  general case $ K \neq \frac12$, an additional contribution to the free energy in the low-temperature phase is expected, which can make the 1-RSB solution stable. The study of these issues is postponed for the future.
 
\textit{7. Acknowledgements.}
We are grateful to V.B. Geshkenbein, A.S. Ioselevich and Y. V. Fyodorov for numerous useful discussions. The work was supported by the RSF grant 20-12-00361, and the grant of the Foundation for the Advancement of Theoretical Physics ``Basis''.

\onecolumngrid

\section{Schwinger-Dyson identities}

The Eq. \eqref{eq:PartitionFunction} from the main paper text is a good starting point for the diagram
technique in terms of auxiliary field $\varphi$. It would then be useful to 
derive exact identities which relate its correlation functions to the correlation
functions of vortex density.

The arbitrary correlation function is defined as follows:

\begin{equation}
	\left\langle O[\delta n,\varphi]\right\rangle \equiv\int{\cal D}\varphi{\rm Tr}_{\text{v}}O[\delta n,\varphi]e^{-S[\varphi,\delta n]}
\end{equation}

Due to the invariance of the integration measure w.r.t. infinitesimal 
transformations $\varphi_{\boldsymbol{r}}^{a}\mapsto\varphi_{\boldsymbol{r}}^{a}+\epsilon_{\boldsymbol{r}}^{a}$,
we immediately  obtain:

\begin{equation}
	\left\langle O[\delta n,\varphi]\right\rangle \equiv\int{\cal D}\varphi{\rm Tr}_{\text{v}}\left(O[\delta n,\varphi]+\sum_{\boldsymbol{r}}\epsilon_{\boldsymbol{r}}^{a}\left[\frac{\partial O[\delta n,\varphi]}{\partial\varphi_{\boldsymbol{r}}^{a}}-O[\delta n,\varphi]\frac{\partial S[\varphi,\delta n]}{\partial\varphi_{\boldsymbol{r}}^{a}}\right]\right)e^{-S[\varphi,\delta n]},
\end{equation}
and due to arbitrary value of $\epsilon_{\boldsymbol{r}}$, taking also 
into account the exact form of the action Eq. (2), we immediately obtain the following
identity:

\begin{equation}
	\left\langle \frac{\partial O[\delta n,\varphi]}{\partial\varphi_{\boldsymbol{r}}^{a}}\right\rangle =\left\langle O[\delta n,\varphi]\frac{\partial S[\varphi,\delta n]}{\partial\varphi_{\boldsymbol{r}}^{a}}\right\rangle =\left\langle O[\delta n,\varphi]\left\{ \sum_{\boldsymbol{r}_{1}}(\beta\hat{J})_{\boldsymbol{r}\boldsymbol{r}_{1}}^{-1}\varphi_{\boldsymbol{r}_{1}}^{a}-i\delta n_{\boldsymbol{r}}^{a}\right\} \right\rangle 
\end{equation}

By picking out various $O$, we can obtain various useful identities for the correlation functions. In particular, we have:

\begin{equation}
	O[\delta n,\varphi]=\varphi_{\boldsymbol{r}^{\prime}}^{b}\Rightarrow\delta_{ab}\delta_{\boldsymbol{r}\boldsymbol{r}^{\prime}}=\sum_{\boldsymbol{r}_{1}}(\beta\hat{J})_{\boldsymbol{r}\boldsymbol{r}_{1}}^{-1}\left\langle \varphi_{\boldsymbol{r}_{1}}^{a}\varphi_{\boldsymbol{r}^{\prime}}^{b}\right\rangle -i\left\langle \delta n_{\boldsymbol{r}}^{a}\varphi_{\boldsymbol{r}^{\prime}}^{b}\right\rangle 
\end{equation}

\begin{equation}
	O[\delta n,\varphi]=i\delta n_{\boldsymbol{r}^{\prime}}^{b}\Rightarrow0=\sum_{\boldsymbol{r}_{1}}(\beta\hat{J})_{\boldsymbol{r}\boldsymbol{r}_{1}}^{-1}i\left\langle \varphi_{\boldsymbol{r}_{1}}^{a}\delta n_{\boldsymbol{r}^{\prime}}^{b}\right\rangle +\left\langle \delta n_{\boldsymbol{r}}^{a}\delta n_{\boldsymbol{r}^{\prime}}^{b}\right\rangle 
\end{equation}
thus the following identity follows:
\begin{equation}
	\left\langle \delta n_{\boldsymbol{r}}^{a}\delta n_{\boldsymbol{r}^{\prime}}^{b}\right\rangle =\delta_{ab}(\beta\hat{J})_{\boldsymbol{r}\boldsymbol{r}^{\prime}}^{-1}-\sum_{\boldsymbol{r}_{1,2}}(\beta\hat{J})_{\boldsymbol{r}\boldsymbol{r}_{1}}^{-1}\left\langle \varphi_{\boldsymbol{r}_{1}}^{a}\varphi_{\boldsymbol{r}_{2}}^{b}\right\rangle (\beta\hat{J})_{\boldsymbol{r}_{2}\boldsymbol{r}^{\prime}}^{-1}
\end{equation}

Furthermore one obtains, for the correlation function  $\left\langle \varphi\varphi\right\rangle $ in the form of Eq. (9):
\begin{equation}
	\left\langle \delta n\delta n\right\rangle =\frac{\hat{{\cal Q}}}{1+\beta\hat{J}\hat{{\cal Q}}}
\end{equation}

It is also worth noting that the local correlation function has then the form $\left\langle \delta n_{\boldsymbol{r}}^{a}\delta n_{\boldsymbol{r}}^{b}\right\rangle =\hat{{\cal Q}}-\hat{{\cal Q}}\hat{{\cal G}}\hat{{\cal Q}}$,
which, strictly speaking, does not coincide with $\hat{{\cal Q}}$; the difference is however parametrically small by an extra $1/W$ factor.

\subsection{Polarizability fluctuations}

The same procedure allows one to obtain the following expression for the  four-point correlation function:

\begin{equation}
	\left\langle \left\langle \delta n_{\boldsymbol{r}_{1}}^{a}\delta n_{\boldsymbol{r}_{2}}^{b}\delta n_{\boldsymbol{r}_{3}}^{c}\delta n_{\boldsymbol{r}_{4}}^{d}\right\rangle \right\rangle =(\beta\hat{J})_{\boldsymbol{r}_{1}\boldsymbol{r}_{1}^{\prime}}^{-1}(\beta\hat{J})_{\boldsymbol{r}_{2}\boldsymbol{r}_{2}^{\prime}}^{-1}(\beta\hat{J})_{\boldsymbol{r}_{3}\boldsymbol{r}_{3}^{\prime}}^{-1}(\beta\hat{J})_{\boldsymbol{r}_{4}\boldsymbol{r}_{4}^{\prime}}^{-1}\left\langle \left\langle \varphi_{\boldsymbol{r}_{1}^{\prime}}^{a}\varphi_{\boldsymbol{r}_{2}^{\prime}}^{b}\varphi_{\boldsymbol{r}_{3}^{\prime}}^{c}\varphi_{\boldsymbol{r}_{4}^{\prime}}^{d}\right\rangle \right\rangle 
\end{equation}

In the vicinity of $T_{c}$, the mean-field theory predicts the following form of the correlation function $\left\langle \left\langle {\cal G}_{\boldsymbol{r}}^{ab}{\cal G}_{\boldsymbol{r}^{\prime}}^{cd}\right\rangle \right\rangle \simeq\left\langle \left\langle \varphi_{\boldsymbol{r}}^{a}\varphi_{\boldsymbol{r}}^{b}\varphi_{\boldsymbol{r}^{\prime}}^{c}\varphi_{\boldsymbol{r}^{\prime}}^{d}\right\rangle \right\rangle$
(for $|\boldsymbol{r}-\boldsymbol{r}^{\prime}|\gg l$, by definition of $\hat{{\cal G}}$ matrix). Using the diagram technique, one can show that coordinate dependence of the $\varphi$ correlation function can be restored in the limit $|\boldsymbol{r}_{1}^{\prime}-\boldsymbol{r}_{2}^{\prime}|\lesssim l$
and $|\boldsymbol{r}_{3}^{\prime}-\boldsymbol{r}_{4}^{\prime}|\lesssim l$ as follows:

\begin{equation}
	\left\langle \left\langle \varphi_{\boldsymbol{r}_{1}^{\prime}}^{a}\varphi_{\boldsymbol{r}_{2}^{\prime}}^{b}\varphi_{\boldsymbol{r}_{3}^{\prime}}^{c}\varphi_{\boldsymbol{r}_{4}^{\prime}}^{d}\right\rangle \right\rangle \simeq G_{\boldsymbol{r}_{1}^{\prime}\boldsymbol{x}}^{aa^{\prime}}G_{\boldsymbol{r}_{2}^{\prime}\boldsymbol{x}}^{bb^{\prime}}G_{\boldsymbol{r}_{3}^{\prime}\boldsymbol{y}}^{cc^{\prime}}G_{\boldsymbol{r}_{4}^{\prime}\boldsymbol{y}}^{dd^{\prime}}\left\langle \left\langle \delta{\cal Q}_{\boldsymbol{x}}^{a^{\prime}b^{\prime}}\delta{\cal Q}_{\boldsymbol{y}}^{c^{\prime}d^{\prime}}\right\rangle \right\rangle 
\end{equation}

Furthermore, since $(\beta\hat{J})_{\boldsymbol{r}\boldsymbol{r}^{\prime}}^{-1}\hat{G}_{\boldsymbol{r}^{\prime}\boldsymbol{x}}=\delta_{\boldsymbol{r}\boldsymbol{x}}-\hat{{\cal Q}}_{\boldsymbol{r}}\hat{G}_{\boldsymbol{r}\boldsymbol{x}}$
and the value of $\hat{{\cal Q}}$ contains an extra smallness  $\sim 1/W$, thus these sections of diagrams can be replaced by delta-functions:

\begin{equation}
	\left\langle \left\langle \delta n_{\boldsymbol{r}}^{a}\delta n_{\boldsymbol{r}}^{b}\delta n_{\boldsymbol{r}^{\prime}}^{c}\delta n_{\boldsymbol{r}^{\prime}}^{d}\right\rangle \right\rangle \simeq\left\langle \left\langle \hat{{\cal Q}}_{\boldsymbol{r}}^{ab}\hat{{\cal Q}}_{\boldsymbol{r}^{\prime}}^{cd}\right\rangle \right\rangle 
\end{equation}

Finally, the mean-square fluctuations of the polarizability corresponds to the replica component $a=c\neq b=d$, which can be symmetrized as follows:

\begin{equation}
	\overline{\left\langle \delta n_{\boldsymbol{r}}\delta n_{\boldsymbol{r}^{\prime}}\right\rangle ^{2}}=\lim_{n\to0}\frac{1}{n(n-1)}\sum_{a\neq b}\left\langle \left\langle \hat{{\cal Q}}_{\boldsymbol{r}}^{ab}\hat{{\cal Q}}_{\boldsymbol{r}^{\prime}}^{ab}\right\rangle \right\rangle ,\quad\lim_{n\to0}\frac{1}{n(n-1)}\mathbb{P}_{bb}^{aa}=\frac{3}{2}
\end{equation}

\section{Derivation of the Ginzburg-Landau functional}

In the main text, the following action for two matrix fields was derived:

\begin{equation}
	nS[\hat{{\cal G}},\hat{{\cal Q}}]=\frac{1}{2}{\rm Tr}(\hat{{\cal G}}\hat{{\cal Q}})+\frac{1}{2}{\rm Tr}\ln(1+\beta\hat{J}\hat{{\cal Q}})+\beta n\sum_{\boldsymbol{r}}F_{\text{v}}[\hat{{\cal G}}_{\boldsymbol{r}}],
\end{equation}
where $F_{\text{v}}[\hat{{\cal G}}]$ is a local free energy of a single-cite problem with the following Hamiltonian:
\begin{equation}
	-\beta\hat{H}_{\text{v}}[\hat{{\cal G}}]=\frac{1}{2}\sum_{ab}\delta n^{a}(\beta^{2}W^{2}+{\cal G}^{ab})\delta n^{b}+\beta\mu\sum_{a}\delta n^{a}
\end{equation}

In this section we will derive the expansion of this action around the 
replica-symmetric solution of saddle point equations in the vicinity of 
the phase transition. Substituting the expansion $\hat{{\cal G}}=\hat{{\cal G}}_{0}+\delta\hat{{\cal G}}$
and $\hat{{\cal Q}}=\hat{{\cal Q}}_{0}+\delta\hat{{\cal Q}}$, the fluctuations of the 
second term can be expressed as follows:

\begin{equation}
	\frac{1}{2}\delta{\rm Tr}\ln(1+\beta\hat{J}\hat{{\cal Q}})=\sum_{k=2}^{\infty}\frac{(-1)^{k+1}}{2k}{\rm Tr}(\hat{G}\delta\hat{{\cal Q}})^{k}=\sum_{k=2}^{\infty}\frac{(-1)^{k+1}}{2k}\frac{{\cal B}_{k}}{a^{2k-2}}{\rm Tr}(\delta\hat{{\cal Q}})^{k}
\end{equation}
(the latter identity utilizes the replicon condition $\sum_{a}\delta{\cal Q}_{ab}=0$), with the following notation:

\begin{equation}
	{\cal B}_{k}=\int(d\boldsymbol{q})G_{0}^{k}(\boldsymbol{q})=\frac{\beta U_{0}}{2}\frac{1}{k-1}\left(\frac{a^{2}}{\nu_{0}T}\right)^{k-1},\quad k>1
\end{equation}

The fluctuations of the third term read:

\begin{equation}
	\beta n\delta F_{\text{v}}[\hat{{\cal G}}]=-\sum_{k=2}^{\infty}\frac{1}{2^{k}k!}Q_{(a_{1}b_{1})\dots(a_{k}b_{k})}\delta{\cal G}_{a_{1}b_{1}}\dots\delta{\cal G}_{a_{k}b_{k}},
\end{equation}
where the following irreducible correlation function with independent variables being pairs $\delta n_{a_{i}}\delta n_{b_{i}}$ was introduced::

\begin{equation}
	Q_{(a_{1}b_{1})\dots(a_{k}b_{k})}\equiv\left\langle \left\langle (\delta n_{a_{1}}\delta n_{b_{1}})\dots(\delta n_{a_{k}}\delta n_{b_{k}})\right\rangle \right\rangle _{\text{v}},
\end{equation}
and the average is performed w.r.t. the Hamiltonian $\hat{H}_{\text{v}}[\hat{{\cal G}}_{0}]$.

The soft mode in this expansion is $\delta\hat{{\cal G}}=\hat{\Psi}$
and $\delta\hat{{\cal Q}}=Q_{22}\hat{\Psi}$. The term ${\rm Tr}\ln$ then reads explicitly:

\begin{equation}
	\frac{1}{2}\delta{\rm Tr}\ln(1+\beta\hat{J}\hat{{\cal Q}})=\nu_{0}T_{c}\sum_{k=2}^{\infty}\left(-\frac{1}{6}\right)^{k-1}\frac{1}{2k(k-1)}{\rm Tr}\hat{\Psi}^{k}=\nu_{0}T_{c}\left(-\frac{1}{24}{\rm Tr}\hat{\Psi}^{2}+\frac{1}{432}{\rm Tr}\hat{\Psi}^{3}-\frac{1}{5184}{\rm Tr}\hat{\Psi}^{4}+\dots\right)
\end{equation}

On the other hand, the $F_{\text{v}}$ term generates terms with different replica structure:

\begin{equation}
	\beta n\delta^{(3)}F_{\text{v}}[\hat{{\cal G}}]=-\frac{1}{12}\left(Q_{33}\sum_{ab}\Psi_{ab}^{3}+2Q_{222}{\rm tr}\hat{\Psi}^{3}\right)=-\nu_{0}T\left(\frac{1}{360}\sum_{ab}\Psi_{ab}^{3}+\frac{1}{180}{\rm tr}\hat{\Psi}^{3}\right)
\end{equation}

\begin{multline}
	\beta n\delta^{(4)}F_{\text{v}}[\hat{{\cal G}}]=-\left(\frac{5}{32}Q_{2222}{\rm tr}\hat{\Psi}^{4}+\frac{1}{48}Q_{44}\sum_{ab}\Psi_{ab}^{4}+\frac{1}{8}Q_{422}\sum_{abc}\Psi_{ab}^{2}\Psi_{ac}^{2}+\frac{1}{4}Q_{332}\sum_{abc}\Psi_{ab}^{2}\Psi_{ac}\Psi_{bc}\right)\\
	=-\nu_{0}T\left(\frac{1}{896}{\rm tr}\hat{\Psi}^{4}+\frac{1}{2016}\sum_{ab}\Psi_{ab}^{4}+\frac{1}{840}\sum_{abc}\Psi_{ab}^{2}\Psi_{ac}\Psi_{bc}-\frac{1}{840}\sum_{abc}\Psi_{ab}^{2}\Psi_{ac}^{2}\right)
\end{multline}
where we have denoted:

\begin{equation}
	Q_{222}=\int\frac{\nu(u)du}{(2\cosh\frac{\beta(u-\mu)}{2})^{6}}\approx\frac{\nu_{0}T}{30},\quad Q_{2222}=\int\frac{\nu(u)du}{(2\cosh\frac{\beta(u-\mu)}{2})^{8}}\approx\frac{\nu_{0}T}{140}
\end{equation}

\begin{equation}
	Q_{33}=\int\frac{\nu(u)du\tanh^{2}\frac{\beta(u-\mu)}{2}}{(2\cosh\frac{\beta(u-\mu)}{2})^{4}}\approx\frac{\nu_{0}T}{30},\quad Q_{332}=\int\frac{\nu(u)du\tanh^{2}\frac{\beta(u-\mu)}{2}}{(2\cosh\frac{\beta(u-\mu)}{2})^{6}}\approx\frac{\nu_{0}T}{210}
\end{equation}

\begin{equation}
	Q_{44}=\int\frac{\nu(u)du\left(\left(2\sinh\frac{\beta(u-\mu)}{2}\right)^{2}-2\right)^{2}}{\left(2\cosh\frac{\beta(u-\mu)}{2}\right)^{8}}\approx\frac{\nu_{0}T}{42},\quad Q_{422}=\int\frac{\nu(u)du\left(\left(2\sinh\frac{\beta(u-\mu)}{2}\right)^{2}-2\right)}{(2\cosh\frac{\beta(u-\mu)}{2})^{8}}=-\frac{\nu_{0}T}{105}
\end{equation}

\section{One-step replica symmetry breaking}

The free energy per lattice cite in the saddle point approximation 
(neglecting the spatial fluctuations of matrices) contains several
terms $\beta F=S[\hat{{\cal G}},\hat{{\cal Q}}]/N=(S_{\text{L}}[\hat{{\cal G}},\hat{{\cal Q}}]+S_{\text{f}}[\hat{{\cal Q}}])/N+\beta F_{\text{v}}[\hat{{\cal G}}]$,
which in the one-step replica symmetry breaking (1RSB) scheme read:

\begin{equation}
	S_{\text{L}}[\hat{{\cal G}},\hat{{\cal Q}}]/N={\rm tr}(\hat{{\cal G}}\hat{{\cal Q}})/2n=\frac{1}{2}\left(-\frac{1-m}{m}{\cal G}_{0}{\cal Q}_{0}+\frac{1}{m}{\cal G}_{0}{\cal Q}_{1}+{\cal G}_{0}{\cal Q}_{2}+{\cal G}_{1}{\cal Q}_{1}+m{\cal G}_{1}{\cal Q}_{2}+{\cal G}_{2}{\cal Q}_{1}\right)
\end{equation}

\begin{equation}
	S_{\text{f}}/N={\rm Tr}\ln(1+\beta\hat{J}\hat{{\cal Q}})/2Nn
	=\frac{\beta U_{0}}{4}\left(-\left(\frac{1}{m}-1\right)\left({\cal Q}_{0}+{\cal Q}_{0}\ln\frac{1}{\beta U_{0}{\cal Q}_{0}}\right)+\frac{1}{m}\left({\cal Q}_{1}+{\cal Q}_{1}\ln\frac{1}{\beta U_{0}{\cal Q}_{1}}\right)+{\cal Q}_{2}\ln\frac{1}{\beta U_{0}{\cal Q}_{1}}\right)
\end{equation}

\begin{equation}
	\beta F_{\text{v}}=\frac{1}{2}({\cal G}_{0}+m{\cal G}_{1})\left(\frac{1}{2}-K\right)^{2}-\frac{1}{8}{\cal G}_{0}-\beta\widetilde{\mu}\left(\frac{1}{2}-K\right)-\frac{1}{m}\int du_{2}\nu_{2}(u_{2})\ln\Xi(u_{2}),
\end{equation}
where we have introduced renormalized chemical potential $\widetilde{\mu}=\mu+T\left({\cal G}_{0}+m{\cal G}_{1}\right)\left(\frac{1}{2}-K\right)$, renormalized disorder strength $\widetilde{W}=\sqrt{W^{2}+T^{2}{\cal G}_{2}}$,
and two auxiliary ``distribution functions'':

\begin{equation}
	\nu_{2}(u_{2})=\frac{\exp(-u_{2}^{2}/2\widetilde{W}^{2})}{\sqrt{2\pi}\widetilde{W}},\quad\nu_{1}(u_{1},u_{2})=\frac{\exp(-u_{1}^{2}/2T^{2}{\cal G}_{1})}{\sqrt{2\pi{\cal G}_{1}}T}\left[2\cosh\frac{\beta(u_{1}+u_{2}-\widetilde{\mu})}{2}\right]^{m},\quad\Xi(u_{2})=\int du_{1}\nu_{1}(u_{1},u_{2})\label{eq:Distributions}
\end{equation}

One can extract the leading asymptotic behavior of the integral over $u_{2}$ 
making use of the small parameter $U_{0}/W \ll 1$:

\begin{equation}
	\beta F_{\text{v}}=\frac{1}{2}({\cal G}_{0}+m{\cal G}_{1})K(1-K)-\beta\widetilde{\mu}\left(\frac{1}{2}-K\right)-\left\langle \ln\left(2\cosh\frac{\beta(u_{2}-\widetilde{\mu})}{2}\right)\right\rangle _{2}-\frac{1}{2}\nu_{0}Tf_{\text{v}}(m,{\cal G}_{1}),
\end{equation}
where the following dimensionless function was introduced:

\begin{equation}
	f_{\text{v}}(m,{\cal G}_{1})=\frac{2}{m}\int dz\left(\ln\Xi(z,m,{\cal G}_{1})-m\ln2\cosh\frac{z}{2}-\frac{m^{2}{\cal G}_{1}}{8}\right)
\end{equation}

\begin{equation}
	\Xi(z,m,{\cal G}_{1})=\Xi\left(u_{2}\equiv\widetilde{\mu}+Tz\right)=\int\frac{dye^{-y^{2}/2{\cal G}_{1}}}{\sqrt{2\pi{\cal G}_{1}}}\left[2\cosh\frac{y+z}{2}\right]^{m},
\end{equation}
with the variables $z=\beta(u_{2}-\widetilde{\mu})$, and $y=\beta u_{1}$.
The variation of the full free energy w.r.t. ${\cal Q}_{i}$ and ${\cal G}_{i}$
yield equations for ${\cal G}_{i}$ and ${\cal Q}_{i}$ respectively --- to Eqs. (\ref{eq:1RSBEquationsG}, \ref{eq:1RSBEquationsQ}).

\subsection{Analysis of the equations in the $T\ll T_{c}$ limit}

The solution in the low temperature limit behaves as $m\ll1$, ${\cal G}_{1}\gg1$, $\xi\equiv m^{2}{\cal G}_{1}/8=O(1)$.
Such scaling allows us to calculate:
\begin{multline}
	\Xi\left(z=\frac{x}{m},m,{\cal G}_{1}\right)\underset{m\ll1}{\equiv}\Xi(x,\xi)=\int\frac{dy}{4\sqrt{\pi\xi}}\exp\left(-\frac{y^{2}}{16\xi}+\frac{1}{2}|y+x|\right)\\
	=\frac{e^{\xi}}{2}\left[e^{y/2}\left(1+\text{erf}\left(\frac{4\xi+x}{4\sqrt{\xi}}\right)\right)+e^{-x/2}\left(1+\text{erf}\left(\frac{4\xi-x}{4\sqrt{\xi}}\right)\right)\right]\label{eq:Xi},
\end{multline}
while for auxiliary dimensionless function the following scaling holds: $f_{\text{v}}(m,{\cal G}_{1})=8f(\xi)/m^{2}$, where:

\begin{equation}
	f(\xi)=\frac{1}{4}\int dx\left(\ln\Xi(x,\xi)-\frac{|x|}{2}-\xi\right)
\end{equation}

The saddle point equations for $q\equiv{\cal Q}_{0}/\nu_{0}T$,
$\xi$ and $m$ can be written as follows:

\begin{equation}
	\begin{cases}
		q & =f^{\prime}(\xi)/(1-m)\\
		\xi & =\frac{3}{4}m\beta T_{c}\ln\frac{1}{f^{\prime}(\xi)}\\
		2f(\xi)-\xi f^{\prime}(\xi) & =\frac{3}{4}m\beta T_{c}(1-q)
	\end{cases}
\end{equation}

Assuming the scaling $m=\mu(T/T_{c})$, $\mu=O(1)$, the system of equations becomes fully dimensionless,
and can be reduced to the single equation for $\xi$ variable, which can then be solved numerically:

\begin{equation}
	\xi=\frac{2f(\xi)-\xi f^{\prime}(\xi)}{1-f^{\prime}(\xi)}\ln\frac{1}{f^{\prime}(\xi)}\Rightarrow\xi\approx9.17
\end{equation}

\begin{equation}
	q=f^{\prime}(\xi)\approx1.43\cdot10^{-5},\quad\mu=\frac{4(2f(\xi)-\xi f^{\prime}(\xi))}{3(1-q)}\approx1.10
\end{equation}

Due to the large value of $\xi$, these numerical solutions can be obtained
analytically with good precision. The scaling function has the following asymptotic
behavior:

\begin{equation}
	f(\xi)\approx\frac{\pi^{2}}{24}-\frac{1}{4}\sqrt{\frac{\pi}{\xi}}e^{-\xi},\quad \xi \gg 1
\end{equation}
so that $q\approx\frac{1}{4}\sqrt{\pi/\xi}e^{-\xi}$ (which yields $1.52\cdot10^{-5}$),
and $\mu\approx8f(\xi)/3\approx\pi^{2}/9$ (which yields $1.10$). Substituting also this
asymptotic to the equation for $\xi$, one can see that it does contain numerically small
parameter $\epsilon=\frac{6}{\pi^{2}}-\frac{1}{2}\approx0.11$ and has the approximate form $\ln\frac{16\xi}{\pi}\approx\epsilon\xi$.

\subsection{Distribution function of the local pinning potential.}

The distribution function of the vortex local pinning potential is defined as follows:

\begin{equation}
	P(u)=\left\langle \left\langle \delta(u-(u_{1}+u_{2})\right\rangle _{1}\right\rangle _{2}\equiv\int du_{2}\nu_{2}(u_{2})\frac{1}{\Xi(u_{2})}\int du_{1}\nu_{1}(u_{1},u_{2})\delta(u-(u_{1}+u_{2})),
\end{equation}
where the averages $\left\langle \dots\right\rangle _{1}$ and $\left\langle \dots\right\rangle _{2}$ are taken w.r.t. distribution functions defined in \eqref{eq:Distributions}.

At low temperatures, the distribution function is noticeably modified in the 
vicinity of the chemical potential in the region of size $\propto T_{c}$. 
We will then calculate the distribution function if the rescaled variable  
$h\equiv(u-\widetilde{\mu})/T_{c}$. The asymptotic behavior of the function 
$\Xi(u_{2})$ was already obtained above, see Eq. \eqref{eq:Xi}, where $z=\beta(u_{2}-\widetilde{\mu})$. 
In this limit, the distribution function reads:

\begin{equation}
	P(h)=\nu_{0}T_{c}\cdot\frac{\exp\left(\mu|h|/2\right)}{4\sqrt{\pi\xi}}\int\frac{dx}{\Xi(x,\xi)}\exp\left(-\frac{(\mu h-x)^{2}}{16\xi}\right)
\end{equation}

Just like in the previous Section, these expression can be further simplified analytically for $\xi\gg1$, and
read as follows:

\begin{equation}
	\Xi(x,\xi)\approx\exp\left(|x|/2+\xi\right),\quad P(h)\approx\nu_{0}T_{c}\cdot\frac{1}{2}\text{erfc}\left(\frac{4\xi-\mu|h|}{4\sqrt{\xi}}\right).
\end{equation}

\subsection{Low temperature behavior of the entropy}

The expression for the entropy can be obtained by differentiating the full free energy w.r.t. $T$.
If one also takes into account the saddle point equations, one obtains the following simple expression valid for arbitrary $T$:

\begin{equation}
	S=\nu_{0}T\left(f_{\text{v}}(m,{\cal G}_{1})+\frac{1}{2}m\frac{\partial f_{\text{v}}}{\partial m}-{\cal G}_{1}\frac{\partial f_{\text{v}}}{\partial{\cal G}_{1}}+\frac{\pi^{2}}{3}\right)-3\beta T_{c}{\cal Q}_{0}
\end{equation}

As we have shown above, in the low temperature limit  auxiliary function
$ f_{\text{v}}$satisfies  scaling relation
$f_{\text{v}}(m,{\cal G}_{1})\approx8f_{\text{v}}(\xi)/m^{2}$. This scaling relation nullifies the combination of 
first three terms in the equation above. However, since $f_{\text{v}}\propto\beta^{2}$, such cancellation 
only guarantees the absence of the unphysical terms $\sim1/T$ in the entropy. In order to extract the 
low-temperature behavior of the entropy, one  should consider corrections to this scaling:

\begin{equation}
	\Delta f_{\text{v}}(m,{\cal G}_{1})\equiv f_{\text{v}}(m,{\cal G}_{1})-\frac{8}{m^{2}}f(\xi)=\frac{2}{m^{2}}\int dx\ln\frac{\Xi(\frac{x}{m},m,{\cal G}_{1})}{\Xi(x,\xi)}-\frac{\pi^{2}}{3}
\end{equation}

The quantity under the logarithm is close to unity when $m \ll 1$, which allows us to expand:

\begin{equation}
	\Delta f_{\text{v}}(m,{\cal G}_{1})=\frac{2}{m^{2}}\int\frac{dx}{\Xi(x,\xi)}\int\frac{dy}{4\sqrt{\pi\xi}}\exp\left(-\frac{y^{2}}{16\xi}\right)\left(\left[2\cosh\frac{y+x}{2m}\right]^{m}-e^{|x+y|/2}\right)-\frac{\pi^{2}}{3}\underset{m\ll1}{\approx}\frac{\pi^{2}}{3}(g(\xi)-1)
\end{equation}
with:

\begin{equation}
	g(\xi)=\frac{1}{4\sqrt{\pi\xi}}\int\frac{dx}{\Xi(x,\xi)}\exp\left(-\frac{x^{2}}{16\xi}\right)\equiv\frac{P(h=0)}{\nu_{0}T_{c}}
\end{equation}
The low-temperature entropy then reads $S=-3\beta T_{c}{\cal Q}_{0}+\frac{\pi^{2}}{3}\nu_{0}Tg(\xi)\to-3\beta T_{c}{\cal Q}_{0}$.


\begin{thebibliography}{40}

\bibitem{NPhys20} B.Sacepe, M.V.Feigel'man and T.M.Klapwijk, \textit{Nat. Phys.}, \textbf{16}, issue 7 (July 2020)
https://doi.org/10.1038/s41567-020-0905-x

\bibitem{Yazdani2013} S. Misra, L. Urban, M. Kim, G. Sambandamurthy, and A. Yazdani
Phys. Rev. Lett. \textbf{110}, 037002 (2013). 

\bibitem{Sacepe2019}  B. Sac{\'e}p{\'e}, J. Seidemann, F. Gay,  K. Davenport, A. Rogachev, M. Ovadia, K. Michaeli, and
M. V. Feigel'man, Nature Physics, \textbf{15}, 48 (2019).

\bibitem{AnnPhys2010}  M.V.Feigel'man, L.B.Ioffe, V.E.Kravtsov and E.Cuevas, Ann.Phys. \textbf{325}, 1390 (2010). 

\bibitem{AAA} A.A. Abrikosov, JETP \textbf{5}, 1174 (1957)

\bibitem{Larkin1970} A.I.Larkin, ZhETF \textbf{58}, 1466 (1970)

\bibitem{LO1979} A. I. Larkin and Yu.N.Ovchinnikov, J.Low Temp.Phys. \textbf{34}, 409 (1979)

\bibitem{PinningReview1} H. Brandt, J.Low Temp.Phys. \textbf{26}, 709 (1977)

\bibitem{PinningReview2} G. Blatter et al., Rev. Mod. Phys. \textbf{66}, 1125 (1994)

\bibitem{PinningReview3}  W.-K. Kwok et al., Rep. Progr. Phys.\textbf{79}, 116501 (2016).

\bibitem{Lab} R.Labusch, Cryst. Lattice Defects \textbf{1}, 1 (1969)

\bibitem{Gesh1} G.Blatter,  V.B.Geshkenbein  and  J.A.G.Koopmann, Phys. Rev. Lett. \textbf{92}, 067009 (2004)

\bibitem{Gesh2} M.Buchacek,  R.Willa, V.B.Geshkenbein  and  G.Blatter, Phys. Rev. B \textbf{98}, 094510 (2018).

\bibitem{ES} A. L. Efros and B. I. Shklovskii, J. Phys. C \textbf{8}, L49 (1975);
A.L.Efros, J. Phys. C \textbf{9}, 2021 (1976)

\bibitem{Nelson1995} U.C.Tauber and D.R.Nelson, Phys. Rev. B \textbf{52}, 16106 (1995).

\bibitem{LarkinKhmelnitsky} A.I.Larkin and D.E.Khmelniskii, ZhETF \textbf{83}, 1140 (1982).

\bibitem{IM2004} M.Mueller and L.B.Ioffe, Phys. Rev. Lett. \textbf{93}, 256403 (2004).

\bibitem{Pankov} S. Pankov and V. Dobrosavljevic, Phys. Rev. Lett. \textbf{94}, 046402 (2005).

\bibitem{MullerPankov} M.Mueller and S.Pankov, Phys.Rev.B \textbf{75}, 144201 (2007)

\bibitem{GrossKanterSompolinsky} D. Gross, I. Kanter and H. Sompolinsky, Phys. Rev.Lett. \textbf{55}, 304 (1985).

\bibitem{LeDoussal2001} David Carpentier, and Pierre Le Doussal, Phys. Rev. E \textbf{63}, 026110 (2001)

\bibitem{MezardBook} M. M{\'e}zard, G. Parisi, and M. A. Virasoro, Spin Glass Theory and Beyond (World Scientific, Singapore, 1987)


\end{thebibliography}
\end{document}